\documentstyle[11pt,aaspp4,flushrt]{article}

\slugcomment{\it to appear in The Astrophysical Journal (July 1, 2000)}
\lefthead{Stern et al.}
\righthead{Cosmological One-Liners }

\def\cf{{c.f.,~}}
\def\ie{{i.e.,~}}
\def\eg{{e.g.,~}}
\def\etal{{et al.~}}

\def\deg{\ifmmode {^{\circ}}\else {$^\circ$}\fi}

\def\secper{\ifmmode \rlap.{^{s}}\else $\rlap{.}{^{s}} $\fi}

\def\kms{\ifmmode {\rm\,km\,s^{-1}}\else
    ${\rm\,km\,s^{-1}}$\fi}
\def\kmsMpc{\ifmmode {\rm\,km\,s^{-1}\,Mpc^{-1}}\else
    ${\rm\,km\,s^{-1}\,Mpc^{-1}}$\fi}
\def\ergcm2s{\ifmmode {\rm\,ergs\,cm^{-2}\,s^{-1}}\else
    ${\rm\,ergs\,cm^{-2}\,s^{-1}}$\fi}
\def\ergsHz{\ifmmode {\rm\,ergs\,s^{-1}\,Hz^{-1}}\else
    ${\rm\,ergs\,s^{-1}\,Hz^{-1}}$\fi}
\def\ergs{\ifmmode {\rm\,ergs\,s^{-1}}\else
    ${\rm\,ergs\,s^{-1}}$\fi}
\def\WHz{\ifmmode {\rm\,W\,Hz^{-1}}\else
    ${\rm\,W\,Hz^{-1}}$\fi}

\def\spose#1{\hbox to 0pt{#1\hss}}
\def\simlt{\mathrel{\spose{\lower 3pt\hbox{$\mathchar"218$}}
     \raise 2.0pt\hbox{$\mathchar"13C$}}}
\def\simgt{\mathrel{\spose{\lower 3pt\hbox{$\mathchar"218$}}
     \raise 2.0pt\hbox{$\mathchar"13E$}}}

\def\lya{Ly$\alpha$}
\def\oxytwo{[\ion{O}{2}]}
\def\magtwo{\ion{Mg}{2}}
\def\hbeta{H$\beta$}
\def\oxythree{[\ion{O}{3}]}
\def\halpha{H$\alpha$}
\def\oiipair{[\ion{O}{2}] $\lambda \lambda 3726,3729$}

\def\woxytwo{W_{\rm [OII]}}
\def\lmm{lines mm$^{-1}$}

\begin{document}

\title{One-Line Redshifts and \\
Searches for High-Redshift \lya\ Emission\altaffilmark{1}}

\author{Daniel Stern\altaffilmark{2}, Andrew Bunker\altaffilmark{3}, Hyron Spinrad}
\affil{Department of Astronomy, University of California at Berkeley \\
Berkeley, CA 94720 }
\and
\author{Arjun Dey\altaffilmark{4}}
\affil{KPNO/NOAO \\
950 N. Cherry Avenue, P.O. Box 26732 \\
Tucson, AZ 85726 }

\altaffiltext{1}{Based on observations at the W.M. Keck Observatory,
which is operated as a scientific partnership among the University of
California, the California Institute of Technology, and the National
Aeronautics and Space Administration.  The Observatory was made possible
by the generous financial support of the W.M. Keck Foundation.}

\altaffiltext{2}{Current address:  Jet Propulsion Laboratory, California
Institute of Technology, Mail Stop 169-327, Pasadena, CA 91109; {\tt
stern@zwolfkinder.jpl.nasa.gov}}

\altaffiltext{3}{Current address:  Institute of Astronomy, Madingley Road,
Cambridge, CB3 OHA, England}

\altaffiltext{4}{Hubble Fellow}

\begin{abstract}

We report the serendipitous discovery of two objects close in
projection with fairly strong emission lines at long wavelength
($\lambda\sim 9190$\,\AA). One (A) seems not to be hosted by any galaxy
brighter than $V_{555}=27.5$, or $I_{814}=26.7$ (Vega-based 3$\sigma$
limits in 1\farcs0 diameter apertures), while the other line is
associated with a faint ($I_{814}\simeq 24.4$) red galaxy (B) offset by
2\farcs7 and 7 \AA\ spectrally.  Both lines are broad (FWHM $\approx
700$\,km\,s$^{-1}$), extended spatially, and have high equivalent
widths ($W_\lambda^{\rm obs}({\rm A}) > 1225$ \AA, 95\% confidence
limit; $W_\lambda^{\rm obs}({\rm B}) \approx 150$ \AA).  No secondary
spectral features are detected for galaxy A.  Blue continuum and the
marginal detection of a second weak line in the spectrum of galaxy B is
consistent with \oxytwo\ (the strong line) and \magtwo\ (the weak line)
at $z = 1.466$.  By association, galaxy A is likely at $z=1.464$,
implying a rest-frame equivalent width of the \oxytwo\ emission line in
excess of 600\,\AA\ and a projected separation of 30~$h_{50}^{-1}$~kpc
for the galaxy pair.  Conventional wisdom states that isolated emission
lines with rest-frame equivalent widths larger than $\sim 200$ \AA\,
are almost exclusively \lya.  This moderate-redshift discovery
therefore compromises recent claims of high-redshift \lya\ emitters for
which other criteria (\ie line profile, associated continuum
decrements) are not reported.  We discuss observational tests to
distinguish \lya\ emitters at high redshift from foreground systems.

\end{abstract}


\section{Introduction}

Serendipitous detections of emission line galaxies are common on
low-dispersion spectrograms taken with large ground-based telescopes.
Indeed, finding distant galaxies through blank sky slit spectroscopy is
fully complementary to narrow-band imaging searches for distant
line-emitting galaxies:  rather than probing a large area of sky for
objects over a limited range of redshift, deep slit spectroscopy
surveys a smaller area of sky for objects at a larger range in redshift
\markcite{Pritchet:94, Thompson:95}(\eg Pritchet 1994; Thompson \&
Djorgovski 1995).  Also, since the resolution of optical spectra is
better matched to narrow line emission than filters with widths of
$\approx 3000$ \kms, deep slit spectroscopy is vastly more sensitive
than narrow-band imaging.  In a 1.5 hour spectrum with the Keck
telescope at moderate dispersion ($\lambda / \Delta \lambda \simeq
1000$), the limiting flux density probed for spectrally unresolved line
emission in a 1 arcsec$^2$ aperture is $S_{\rm lim} (3 \sigma) \approx
6 \times 10^{-18} \ergcm2s$ at $\lambda \approx 9300$ \AA.  The first
galaxy confirmed at $z > 5$ was found serendipitously during
spectroscopic observations of a galaxy at $z = 4.02$
\markcite{Dey:98}(Dey {et~al.} 1998).  Since then, several other $z >
5$ galaxies have been reported \markcite{Hu:98, Weymann:98, Spinrad:98,
vanBreugel:99a, Chen:99, Hu:99}(Hu, Cowie, \& McMahon 1998; Weymann
{et~al.} 1998; Spinrad {et~al.} 1998; van Breugel {et~al.} 1999; Chen,
Lanzetta, \& Pascarelle 1999; Hu, McMahon, \& Cowie 1999).

In both narrow-band imaging and slit spectroscopy surveys, the
equivalent width of detected emission lines is a standard redshift
indicator \markcite{Cowie:98, Hu:98}(\eg Cowie \& Hu 1998; Hu {et~al.}
1998).  \lya\ at high redshift can have a very large rest-frame
equivalent width \markcite{Charlot:93}(up to $W_\lambda^{\rm rest}
\approx 200$ \AA\ if driven by star formation; Charlot \& Fall 1993),
while \oxytwo $\lambda$3727\AA\ at low to moderate redshift, the other
primary strong, solitary emission feature in UV/optical galaxy spectra,
rarely has a rest-frame equivalent width exceeding 100 \AA\, in
magnitude-limited surveys \markcite{Songaila:94, Guzman:97, Hammer:97,
Hogg:98}(\eg Songaila {et~al.} 1994; Guzm\`an {et~al.} 1997; Hammer
{et~al.} 1997; Hogg {et~al.} 1998).  Equivalent width selection is also
helped by $W_\lambda^{\rm obs} = W_\lambda^{\rm rest} (1 + z)$, which
boosts \lya\ more than \oxytwo.  The other strong UV/optical emission
features in galaxies, \eg \hbeta,
\oxythree$\lambda\lambda$4959,5007\AA, \halpha, are generally easily
identified spectroscopically from their wavelength proximity to other
emission features, though \halpha\ can also have extremely high
equivalent widths, up to 3000 \AA\ \markcite{Leitherer:95}(Leitherer,
Carmelle, \&  Heckman 1995), leading to some confusion between
low-redshift, young, dwarf starbursts and high-redshift \lya-emitters
\markcite{Stockton:98}(\eg Stockton \& Ridgway 1998).

Here we report the discovery of two emission lines in the approximate
direction of the Abell~2390 cluster.  The equivalent widths of these
lines are rather large, with $W_\lambda^{\rm obs} (\rm A) > 1225$
\AA\ (95\% confidence limit) for the first object and $W_\lambda^{\rm
obs} (\rm B) \approx 150$ \AA\, for the second object.  However, as we
argue below, \oxytwo\ at $z=1.46$ is the most likely identification for
these features, indicating an atypical system.

In \S 2 $-$ \S 4 we discuss the observations, redshift determination,
and properties of these galaxies.  In \S 5 we consider the implications
of this discovery for narrow-band searches for \lya\ emission from
distant protogalaxies, including a detailed discussion of the
observational criteria useful for distinguishing high-redshift
\lya\ from foreground emission-line galaxies.  

Throughout this paper, unless otherwise indicated, we adopt $H_0 = 50~
h_{50}~ \kmsMpc, q_0 = 0.1$, and $\Lambda = 0$.  For these parameters,
the luminosity distance, $d_L$, at $z=1.46$ is 13.88 $h_{50}^{-1}$ Gpc
and 1\arcsec\ subtends 11.1 $h_{50}^{-1}$ kpc.


\begin{figure}[t!]
\plotfiddle{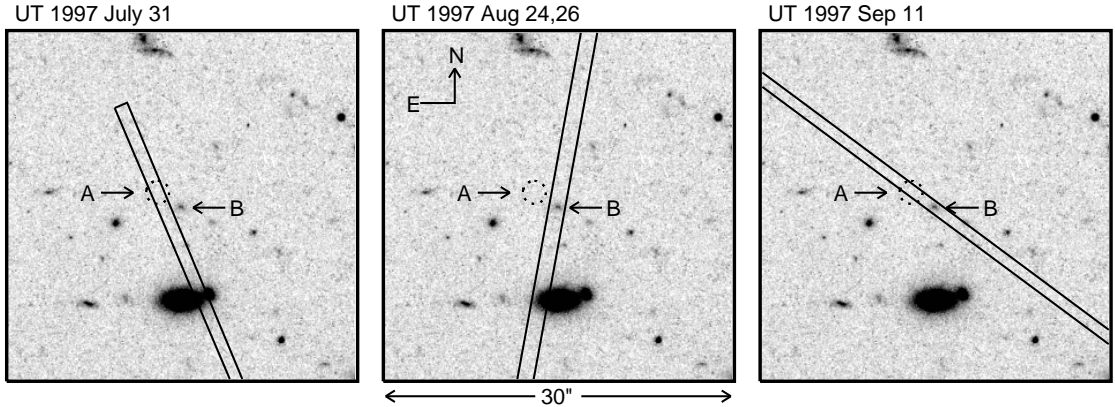}{2.0in}{0}{91}{91}{-220}{0}

\caption{{\it HST} $I$-band (F814W; 10500 s) image of the
serendipitously discovered galaxies in the field of Abell~2390.  Galaxy
A, marked by dotted circle, is undetected ($I_{814} > 26.7, 3\sigma$).
Galaxy B has $I_{814} = 24.4$ (1\arcsec\ diameter aperture).  Slit
positions for our three observations are indicated.  The images shown
are 30\arcsec\ on a side, oriented with north upwards and east to the
left.  Galaxy B is located at $\alpha = 21^h53^m35\secper16$, $\delta =
+17\deg42\arcmin56\farcs66$ (J2000) and is 2\farcs7 SE of galaxy A.
The bright elliptical galaxy south of galaxy B is a member of the Abell
2390 galaxy cluster ($z=0.241$).}

\label{hstslit}
\end{figure}

\section{Observations}

On UT 1997 July 31, an observation of a pair of lensed arcs close to
the core of the rich galaxy cluster Abell~2390 \markcite{Frye:98,
Bunker:00a}(Frye \& Broadhurst 1998; Bunker, Moustakas, \& Davis 2000) resulted in the serendipitous detection of a strong emission
line at 9185 \AA\ approximately 70\arcsec\ NNE of the arcs along the
long slit spectrogram.  These observations were made with the Low
Resolution Imaging Spectrometer \markcite{Oke:95}(LRIS; Oke {et~al.} 1995) at the
Cassegrain focus of the Keck~II Telescope, using the 400
\lmm\ grating ($\lambda_{\rm blaze} \approx 8500$ \AA).  The detector
is a Tek $2048^2$ CCD with $24 \mu$m pixels, corresponding to
0\farcs212 pix$^{-1}$.  The GG495 filter was used to block second
order light and the observations sample $\lambda\lambda 6100 - 9900$
\AA.  The slit width was 1\arcsec, yielding an effective resolution of
$\Delta \lambda_{\rm FWHM} \approx 8$ \AA, and the data were binned by
a factor of two spatially during the data acquisition.  Reductions were
done with IRAF and followed standard spectroscopic procedures.
Wavelength calibration was verified against telluric emission lines.
The night was photometric with a seeing of $\approx$ 0\farcs7 FWHM, and the
data were calibrated using observations of BD+174708 \markcite{Massey:88,
Massey:90}(Massey {et~al.} 1988; Massey \& Gronwall 1990) procured on the same night.  The total integration time was
3600 s with the slit oriented at a position angle of $23$\deg\ (see
Fig.~1).  The emission line source (Fig.~2), which we refer to as
galaxy A, was spatially extended by 3\arcsec\ in this discovery
spectrogram and no associated continuum was detected, implying an
equivalent width $W_\lambda^{\rm obs} > 400$ \AA\ (95\% confidence
limit).  The wavelength of the emission line does not correspond to any
prominent lines at the redshift of Abell~2390 ($z = 0.24$).  Throughout we have
corrected for foreground Galactic extinction using the dust maps of
\markcite{Schlegel:98}Schlegel, Finkbeiner, \& Davis (1998) which have an optical reddening of $E(B-V) = 0.11$
in the direction of Abell~2390, equivalent to extinctions of $A_{555} =
0.33$ and $A_{814} = 0.20$.


\begin{figure}[t!]
\plotfiddle{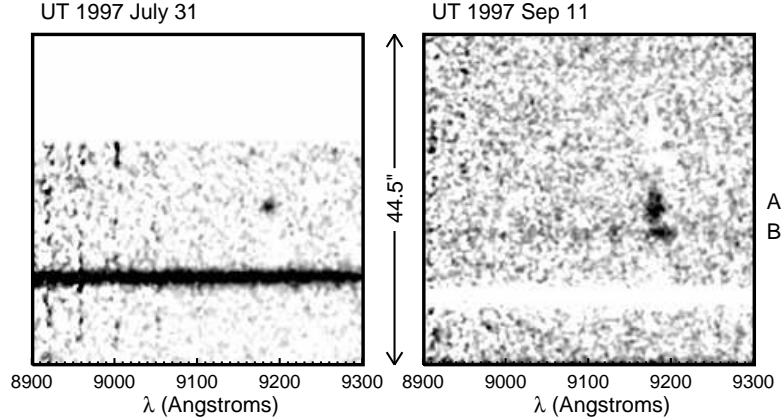}{2.0in}{0}{91}{91}{-160}{0}

\caption{Two-dimensional, flattened, sky-subtracted spectra near the
emission lines from our discovery (July; left) and final (September;
right) observations.  The abscissa indicates wavelength and the
ordinate indicates spatial position.  The July data do not sample the
entire spatial range.  Both data sets have been convolved with a
Gaussian of width 1 pixel.  Galaxy~A is detected in both observations
and is the upper emission feature in the September data.  Galaxy~B is
only visible in the September data (lower emission feature).  Note the
lack of continuum associated with galaxy~A in both data sets, while
galaxy~B has faint continuum blue-ward of the emission line, indicating
that the emission line is unlikely to be \lya.  The horizontal feature
in the July data set is continuum emission from an Abell~2390 cluster
member ($z=0.241$).  The negative horizontal feature in the September
data is an artifact of our fringe suppression algorithm (see text); it
is the negative of a different Abell~2390 cluster member ($z=0.232$).
Faint negatives of galaxies A and B are also evident.}

\label{twod}
\end{figure}

A comparison with archival {\it Hubble Space Telescope} ({\it HST}) Wide
Field/Planetary Camera~2 \markcite{Trauger:94}(WFPC2; Trauger {et~al.} 1994) images secured
by Fort \etal (HST-GO~5352) finds no obvious optical identification for
object A.  The {\it HST} imaging was undertaken on UT 1994 December 10 in
the F555W ($V_{555}$) and F814W ($I_{814}$) filters.  The data comprised
four orbits in F555W and five in F814W, with each orbit consisting of
a single integration of 2100 s.  For the combined F555W image (8400 s),
the 1$\sigma$ limiting magnitude reached in 1 square arcsecond is $V_{555}
= 28.4$ mag arcsec$^{-2}$, and for the combined F814W image (10500 s),
it is $I_{814} = 27.6$ mag arcsec$^{-2}$ (Vega-based magnitudes are
adopted throughout).  This is consistent with the predicted sensitivity
based on the Poissonian counting statistics and the WF readout noise
(5 $e^-$).  The 3$\sigma$ limits on the brightness of the host of
galaxy A are $V_{555} > 27.5$ and $I_{814} > 26.7$ for a 1\arcsec\
diameter aperture.  A faint red galaxy, which we refer to as galaxy
B, was found nearby ($\approx$ 2\farcs7 to the SE) but {\em not} coincident
with our estimate of the galaxy A position based on our spectroscopic
slit observations (see Fig.~1).  Galaxy B is faintly detected in F555W
($V_{555} = 27.63 \pm 0.54$) and has $I_{814} = 24.36 \pm 0.05$, where
both magnitudes are quoted for 1\arcsec\ diameter apertures.

On UT 1997 August 24 \& 26 we observed galaxy B using LRIS in slitmask
mode at a position angle of $-$10.4\deg\ (Fig.~1).  The
1.4\arcsec\ wide slitlet was 42\arcsec\, long, and observations were
made with both the 600 \lmm\ grating (UT 1997 August 24; 3600 s;
$\lambda_{\rm blaze} = 5000$\AA; $\Delta\lambda_{\rm FWHM} \approx
7.5$\AA; $\lambda\lambda 7050 - 9550$\AA) and the 300 \lmm\ grating (UT
1997 August 26; 4800 s; $\lambda_{\rm blaze} = 5000$\AA;
$\Delta\lambda_{\rm FWHM} \approx 14$\AA; $\lambda\lambda 3800 -
8800$\AA).  Moderate cirrus affected these observations.  Our
spectrogram of galaxy B shows a {\em weak} emission line at $\lambda \simeq 9196$
\AA, nearly the same wavelength as for galaxy A.  However, galaxy B
also shows weak continuum blue-ward of the emission line extending down
to wavelengths of $\sim 4000$\AA.


\begin{figure}[t!]
\plotfiddle{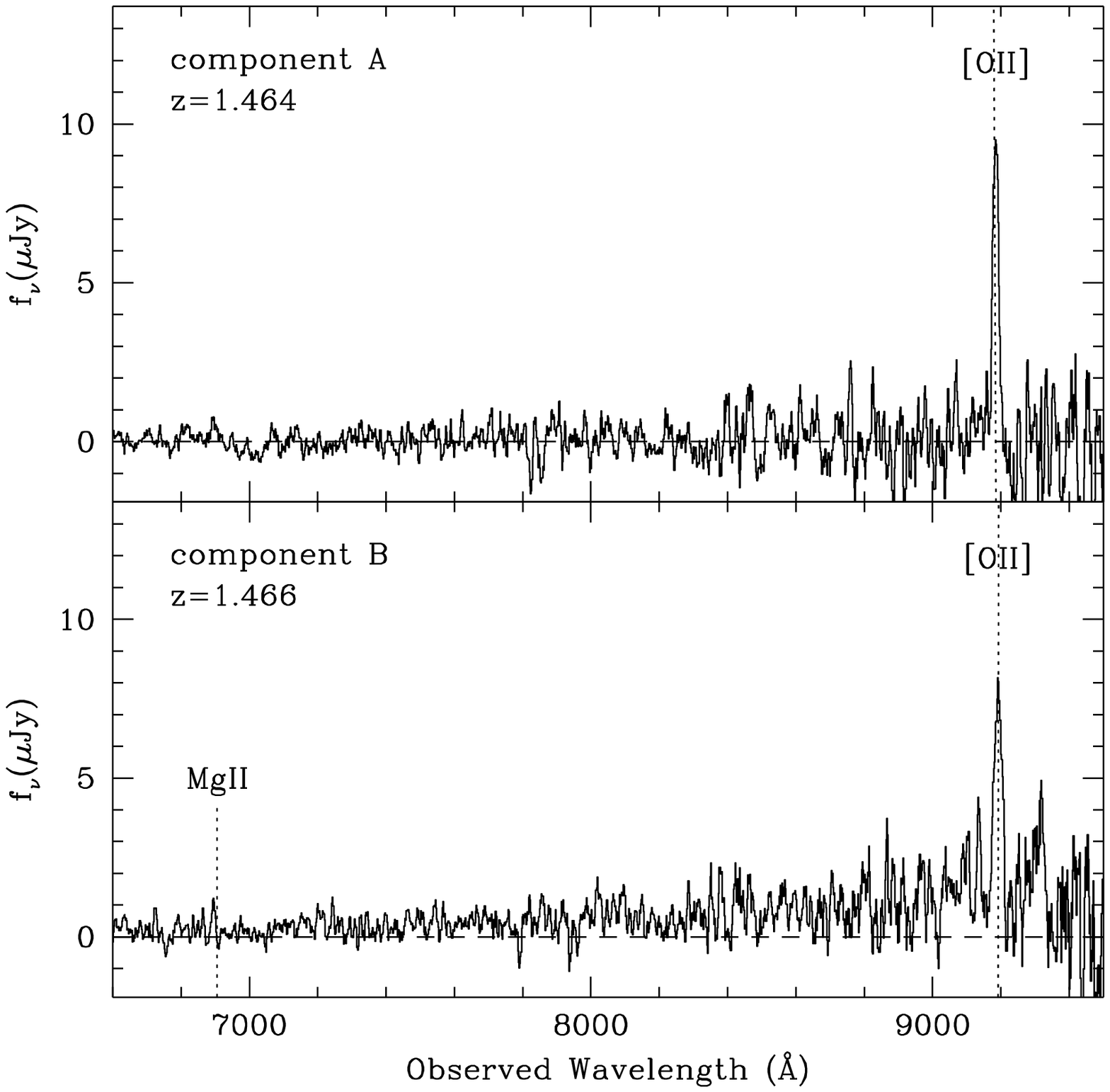}{4.8in}{0}{75}{75}{-240}{-130}

\caption{Spectra of galaxies A and B in the Abell~2390 field from the
UT 1997 Sep 11 observations, illustrating the high equivalent width
emission lines observed at 9185 \AA\ (galaxy A) and 9191 \AA\ (galaxy
B).  The total exposure time was 5400 s, and the spectrum was extracted
using a 1\farcs5 $\times$ 1\farcs5 aperture.}

\label{oned}
\end{figure}

Finally, on UT 1997 September 11, we attempted to put the LRIS 1\farcs0
wide slit across both galaxies, as well as could be determined from
their spectroscopic positions (Fig.~1).  These observations were
obtained with the 400 \lmm\ grating ($\lambda_{\rm blaze} = 8500$ \AA)
at a position angle of 80\deg\ and sample $\lambda\lambda 5950 - 9730$
\AA.  The seeing was $\approx 0\farcs7$.  As the night was not
photometric, we applied the UT 1997 July 31 sensitivity function,
scaled to maintain the same total flux in the galaxy A emission line.
The total integration time was 5400 s.  In Fig.~2 we present the July
and September two-dimensional, flattened, sky-subtracted spectra near
the emission lines.  In order to suppress the fringing which adversely
affects long wavelength ($\lambda \simgt 7200$ \AA) optical spectra, we
dithered the telescope between individual exposures and subtract
temporally adjacent frames prior to co-adding the data sets.  This
procedure, which is essential for recovering information on faint
objects in the telluric OH bands, should not affect the object spectra,
but does leave negative holes in the final two-dimensional spectra (see
Fig.~2).  Fig.~3 presents the extracted, one-dimensional spectra from
the September data and Table~1 summarizes the observed properties of
galaxies A and B.

\begin{deluxetable}{lrcrr}
\tablewidth{0pt}
\tablecaption{Observed Properties of Galaxies A and B in the Abell~2390 Field.}
\tablehead{
\colhead{Parameter} &
\colhead{Galaxy A} &
\colhead{~~~~~~} &
\colhead{} &
\colhead{Galaxy B}}
\startdata
Line ID \dotfill & \oxytwo & & \magtwo & \oxytwo \nl
$\lambda_{\rm obs}$ (\AA) \dotfill & $9184.8\pm0.8$ & & $\simeq$ 6895 & $9191.3
\pm1.4$ \nl
$z$ \dotfill & 1.464 & & 1.463 & 1.466 \nl
$f_{\rm line}$ ($10^{-17}$ ergs~ cm$^{-2}$ s$^{-1}$) \dotfill & $8.7\pm0.8$ & &
0.9 & $7.3\pm0.9$ \nl
$L_{\rm [OII]}$ ($h_{50}^{-2}~ 10^{42}$ ergs~ s$^{-1}$) \dotfill & $2.02 \pm 0.2
$ & & & $1.70 \pm 0.2$ \nl
$f_{\rm cont}$ ($10^{-20}$ ergs~ cm$^{-2}$ s$^{-1}$ \AA$^{-1}$) \dotfill & $-2.2
\pm4.2$ & & & $51.4\pm4.5$ \nl
$W_\lambda^{\rm obs}$ (\AA) \dotfill & $> 1225$ & & 90 & $102 - 189$ \nl
FWHM (km s$^{-1}$) \dotfill & $627\pm62$ & & 300 & $785\pm94$ \nl
$V_{555}$ (mag) \dotfill & $>27.5$ & & & $27.63 \pm 0.54$ \nl
$I_{814}$ (mag) \dotfill & $>26.7$ & & & $24.36 \pm 0.05$ \nl
\enddata
\tablecomments{Magnitudes are in the Vega system and quoted for
1\farcs0 circular apertures.  $3 \sigma$ limits are presented for the
magnitude of galaxy A.  Spectroscopic measurements derive from UT 1997
September 11 data and have been corrected for Galactic extinction using
an optical reddening of $E(B-V) = 0.11$ \markcite{Schlegel:98}(Schlegel
{et~al.} 1998).  The \magtwo\ line is weakly detected.  Parameters for
the \oxytwo\ lines have been measured using the SPECFIT contributed
package within IRAF \markcite{Kriss:94}(Kriss 1994) with a flat (in
$F_\lambda$) continuum and a single Gaussian emission line.  Equivalent
widths have been calculated with a Monte Carlo analysis using the
measured line flux and continuum amplitude with errors subject to the
constraint that both must be $> 0$.  We quote the 95\% confidence limit
for galaxy A and the 95\% confidence interval for galaxy B.  Line
widths (FWHM) are deconvolved by the instrumental resolution ($\approx$
8 \AA).}

\label{specprop}
\end{deluxetable}

\section{Redshift Determination }

Understanding objects A and B requires first determining their
redshifts. Speculations on one-line redshifts are common among faint
galaxy observers.  Here we have been spared that fate by the presence
of the second {\em weak} emission line in the spectrum of object~B at
$\lambda \simeq 6895$ \AA.  The wavelength ratio with the stronger line
at $\lambda = 9191$ \AA\ is 1.333, close to the laboratory measured
[\ion{O}{2}]$\lambda3727$\AA\ /
\ion{Mg}{2}$\lambda\lambda$2796,2803\AA\ = 1.332.  This implies a
redshift $z = 1.466$ for galaxy B with the stronger line identified, as
is usually the case, with [\ion{O}{2}]$\lambda$3727\AA\ (hereinafter
\oxytwo).  For non-active galaxies it is slightly unusual to observe
\ion{Mg}{2}$\lambda\lambda$2796,2803\AA\ (hereinafter \magtwo) in
emission.  For example, \markcite{Guzman:97}Guzm\`an {et~al.} (1997) report on a spectroscopic
study of 51 compact field galaxies in the flanking fields of the Hubble
Deep Field.  Of the 9 galaxies at redshifts sufficient for \magtwo\ to
be sampled by their observations, only 2 (22\%) show \magtwo\ in {\em
emission}.  Active galaxies, of course, often show \magtwo\ in
emission.  The velocity offset between the \magtwo\ and \oxytwo\ lines
are cause for slight concern, but similar offsets are common in radio
galaxy spectra \markcite{Stern:99a}(\eg Stern {et~al.} 1999b).  As discussed below, the
continuum blueward of the long wavelength, isolated line
makes alternate redshift identifications unconvincing.

The question then remains:  what is the redshift of object A?  Previous
experience might suggest that isolated optical emission lines with
$W_\lambda^{\rm obs} > $ several $\times 100$ \AA\ are exclusively
identified with \lya\ at high redshift.  For example, 0140+326~RD1 at
$z=5.34$ has $W_{Ly\alpha}^{\rm obs} = 600 \pm 100$\AA\ \markcite{Dey:98}(Dey {et~al.} 1998)
while HDF~4-473.0 at $z=5.60$ has $W_{Ly\alpha}^{\rm obs} \approx
300$\AA\ \markcite{Weymann:98}(Weymann {et~al.} 1998).  Magnitude-limited redshift surveys find
the $W_\lambda^{\rm obs}$ distribution for \oxytwo, typically the
primary doppelg\"anger for high redshift \lya, rarely has a rest-frame
equivalent width exceeding 100\AA\ \markcite{Songaila:94,
Guzman:97, Hammer:97, Hogg:98}(\eg Songaila {et~al.} 1994; Guzm\`an {et~al.} 1997; Hammer {et~al.} 1997; Hogg {et~al.} 1998).  High values of $\woxytwo$ are
occasionally seen in AGN, however.  The $3^{rd}$ Cambridge/Molonglo
Radio Catalog (3C/MRC) composite radio galaxy spectrum of
\markcite{McCarthy:93}McCarthy (1993) has $\woxytwo = 128$\AA\ while the lower radio
power MIT-Green Bank (MG) composite radio galaxy spectrum of
\markcite{Stern:99a}Stern {et~al.} (1999b) has $\woxytwo = 142$\AA.  Since the surface density
of luminous active galaxies is relatively low, 
our July data suggested at first that the
emission line in object A was enticingly identified
with \lya\ at the extremely high redshift of $z = 6.55$.  Although this
interpretation cannot be {\em completely} ruled out, the
robust identification of the emission line in object B 
with \oxytwo\ at $z = 1.466$ coupled with its projected proximity
to object A strongly argues that we are witnessing associated galaxies
at moderate redshift.  The projected separation at $z = 1.466$ is
$30~ h_{50}^{-1}$ kpc.  The radial velocity difference between the
objects is 220 km s$^{-1}$.


\section{Galaxies A and B as Active Galaxies}

Associating the \oxytwo\ emission in this system with recent star
formation activity is likely inappropriate.  The \oxytwo\ luminosities
imply star formation rates of $\approx 90~h_{50}^{-2}~M_\odot~{\rm
yr}^{-1}$ \markcite{Kennicutt:92}(Kennicutt 1992) which seems improbable given the
faintness of the hosts.  Using the SPECFIT contributed package within
IRAF \markcite{Kriss:94}(Kriss 1994) to fit the emission lines with the
\oiipair\ doublet, we derive deconvolved circular velocities $v_c
\simgt 200 / \sin i$ km s$^{-1}$ for both galaxies.  Applying the
high-redshift Tully-Fisher relation \markcite{Vogt:97}(Vogt {et~al.} 1997), these line widths
suggest apparent $I$-band magnitudes $I \simlt 23.4$, much more
luminous than the observed galaxies.

A more natural explanation for the spectral character of this system is
to associate the line emission from galaxies A and B with active
galactic nuclei (AGN).  The line widths are consistent with those seen
in radio galaxies and Seyfert galaxies:  the deconvolved FWHM of
galaxies A and B are $\approx 650$ and $\approx 800$ km s$^{-1}$
respectively (fit as a single line), while the composite radio galaxy
spectra have FWHM$_{\rm [OII]} \simgt 1000$ km s$^{-1}$
\markcite{McCarthy:99a, Stern:99a}(McCarthy \& Lawrence 1999; Stern {et~al.} 1999b).  This interpretation also presents a
natural explanation for the high equivalent widths, similar to those
seen in other active systems.

Comparison with the FIRST radio catalog \markcite{Becker:95}(Faint Images of the
Radio Sky at Twenty-one cm; Becker, White, \& Helfand 1995) reveals no radio source
within 1 arcminute of either galaxy to a limiting flux density of
$f_{\rm 1.4 GHz} \simeq 1$ mJy ($5 \sigma$).  The traditional
demarcation between radio-loud and radio-quiet systems is $\log L_{\rm
1.4 GHz} ({\rm ergs\ s}^{-1} {\rm Hz}^{-1}) = 32.5$.  For an emitted
luminosity density $L_\nu \propto \nu^\alpha$, this demarcation
corresponds to $S_{\rm 1.4 GHz} = 1.36 h_{50}^2 (1 + z)^{1 + \alpha}$
mJy for $z = 1.46$.  The FIRST non-detection therefore
does not preclude a weak radio-loud source.

\section{Tests for Cosmological One-liners}

Several programs are currently underway to search for high-redshift
primeval galaxies through deep narrow-band imaging
\markcite{Stern:99e}(\eg see Stern \& Spinrad 1999).  Table~2 lists
several recently discovered high-redshift ($z > 5$) sources and
includes lower-redshift, strong line-emitters.  The previous generation
of narrow-band surveys failed to confirm any field \lya-emitting
protogalaxy candidates \markcite{Pritchet:94, Thompson:95}(Pritchet
1994; Thompson \& Djorgovski 1995).  Examples of new programs include
the Calar Alto Deep Imaging Survey \markcite{Thommes:98,
Thommes:99}(CADIS, Thommes {et~al.} 1998; Thommes 1999) which uses a
Fabry-P\'erot interferometer ($S_{\rm lim} (5 \sigma) \approx 3 \times
10^{-17} \ergcm2s $), the Keck-based narrow-band, interference filter
imaging program of Cowie, Hu, and McMahon \markcite{Cowie:98, Hu:98}
(Cowie \& Hu 1998; Hu, Cowie, \& McMahon 1999; $S_{\rm lim}
(5 \sigma) \approx 1.5 \times 10^{-17} \ergcm2s$),
and serendipitous searches on deep slit spectra
\markcite{Manning:00}(\eg this paper; Manning {et~al.} 2000).

\begin{deluxetable}{lccccl}
\tablewidth{0pt}
\tablecaption{Galaxies Reported/Considered at $z > 5$.}
\tablehead{
\colhead{Galaxy} &
\colhead{$z$} &
\colhead{$W_\lambda^{\rm obs}$} &
\colhead{asymmetric} &
\colhead{continuum} &
\colhead{Reference} \nl
\colhead{} &
\colhead{} &
\colhead{(\AA)} &
\colhead{line profile} &
\colhead{decrement} &
\colhead{}}
\startdata
STIS parallel gxyA & 6.68 & $\approx 120$ & \nodata & x & \markcite{Chen:99}Chen
 {et~al.} (1999) \nl
SSA22-HCM1      & 5.74 & 350 & x & x & \markcite{Hu:99}Hu {et~al.} (1999) \nl
BR 1202$-$0725 ser & 5.64 & $>600$ & \nodata & x & \markcite{Hu:98}Hu {et~al.} 
(1998) \nl
HDF 4-473.0     & 5.60 & $\approx 300$ & x & xx & \markcite{Weymann:98}Weymann 
{et~al.} (1998) \nl
0140+326 RD1    & 5.34 & $600\pm100$ & x & x & \markcite{Dey:98}Dey {et~al.} 
1998) \nl
HDF 3-951.0     & 5.34 & N/A & N/A & xx & \markcite{Spinrad:98}Spinrad {et~al.}
(1998) \nl
TN J0924$-$2201 & 5.19 & $\approx 150$ & 1/2 & \nodata & \markcite{vanBreugel:99a}
van Breugel {et~al.} (1999)
\nl
\nl
Abell 2390 serA & 1.46 & $>1225$ & no & \nodata & this paper \nl
ERO~J164502+4626.4 & 1.44 & 115 & \nodata & x & \markcite{Dey:99a}Dey {et~al.} 
(1999) \nl
3C212 B08       & 0.31 & $\approx 640$ & \nodata & no & \markcite{Stockton:98}
Stockton \& Ridgway (1998) \nl
\enddata

\tablecomments{Symbols indicate reliability of the criterion
considered.  Ellipses (`...') refer to unreported criteria.  `x' refers
to a positive result and `xx' refers to an extremely positive result
--- namely, the two $z > 5$ confirmed galaxies in the HDF whose
spectral energy distributions are confirmed flat (in $f_\nu$) into the
near-infrared.  The source HDF 3-951.0 lacks \lya\ emission; we
therefore list some of its criteria not applicable, `N/A'.  The `1/2'
for the high-redshift radio galaxy TN J0924$-$2201 refers to the
enigmatic result that one observation illustrates asymmetric
\lya\ emission while a second observation does not.}

\label{comp5}
\end{deluxetable}

Selecting objects on the basis of strong line emission may sample a
different galaxy population from the traditional magnitude-limited
surveys.  In particular, emission-line surveys are much more sensitive
to active galaxies and objects undergoing massive bursts of star
formation.   Determining the redshift and physical origin of line
emission is challenging; comparison to field surveys selected on the
basis of continuum magnitude is perhaps inappropriate.  Samples of
line-emitting protogalaxy candidates will shortly become available.  We
therefore present below a timely and detailed discussion of the
observational criteria which can be used to distinguish high-redshift
\lya\ emission from low-redshift interlopers.

\subsection{Equivalent Width}

The stellar population synthesis models of \markcite{Charlot:93}Charlot \& Fall (1993) predict
rest-frame \lya\ equivalent widths of $50 - 200$ \AA\ for dust-free
young galaxies.  For a constant star formation history, the \lya\
luminosity and equivalent width are only somewhat dependent on the star
formation rate and are greatest at times less than 10 Myr after the
onset of the burst.  For comparison, the spectral atlas of nearby
galaxies by \markcite{Kennicutt:92}Kennicutt (1992) shows that the rest-frame equivalent
width of the H$\alpha$ + [\ion{N}{2}] complex rarely exceeds 200 \AA,
that of [\ion{O}{3}]$\lambda$5007 \AA\ rarely exceeds 100 \AA, H$\beta$
rarely exceeds 30 \AA, and [\ion{O}{2}] rarely exceeds 100 \AA.  With
the $(1 + z)$ amplification of observed equivalent widths, emission
lines in the optical with measured equivalent widths larger than
$\approx 300$ \AA\ should be almost exclusively identified with \lya.
See Table~2 for a list of the equivalent widths of several recently
reported protogalaxy candidates at $z > 5$.

A number of caveats temper sole reliance on equivalent width arguments
to discriminate \lya\ from foreground emission.  First, high-ionization
\ion{H}{2} dwarf galaxies can have strong H$\alpha$ emission with very
weak continuum.  \markcite{Stockton:98}Stockton \& Ridgway (1998) report an object (3C212 B08) with
a single strong emission line at 8567 \AA\ and $W_\lambda^{\rm obs}
\approx 640$ \AA.  They eventually identify the line as H$\alpha$ at $z =
0.305$ due to a secondary feature at 2\% of the strong line whose
wavelength matches redshifted \ion{He}{1}$\lambda$5876 \AA.  Starburst
models \markcite{Leitherer:95}(Leitherer {et~al.} 1995) with continuous star formation and a
Salpeter initial mass function over the mass range $0.1 - 100$
M$_\odot$ can have H$\alpha$ equivalent widths as high as 3000 \AA\ up
to ages of 3 Myr and can remain above $\sim 300$ \AA\ up to ages of 100
Myr.  Furthermore, extreme \ion{H}{2} galaxies with very hot ($>$
60,000 K) stars and low metal abundances can have suppressed
low-ionization metallic emission lines such as [\ion{N}{2}] and
[\ion{S}{2}], making H$\alpha$ identification difficult \markcite{Terlevich:91}(\eg Tol
1214$-$277; Fig.~4o in Terlevich {et~al.} 1991).  Such galaxies tend to show
\ion{He}{1}$\lambda$5876 \AA\ in emission.

AGN offer another potential source of ionizing radiation to stimulate
line emission:  photoionization by a power law continuum emitted near
the central engine combined with shock-excited emission can produce
very high equivalent width emission.  The composite radio galaxy
spectrum in \markcite{McCarthy:93}McCarthy (1993) shows several lines with rest-frame
equivalent widths in excess of 50 \AA.  As argued earlier, however, the
spectral proximity of several features make the most likely
identification of an observed, isolated, high-equivalent width emission
feature an ambiguous selection between \lya\ and \oxytwo.  Composite
radio galaxy spectra have $\woxytwo \approx 135$
\AA\ \markcite{McCarthy:93, Stern:99a}(McCarthy 1993; Stern {et~al.} 1999b).  Occasional sources show extremely
high equivalent width \oxytwo.  For example, \markcite{McCarthy:91}McCarthy (1991)
reports \oxytwo\ emission from the radio galaxy B3~0903+428 ($z =
0.907$) with $\woxytwo = 251 \pm 55$ \AA\ (rest-frame) and the northern
knot in 3C368 has a rest-frame $\woxytwo = 485$ \AA\ \markcite{Dey:99c}(Dey 1999).
Luminous, narrow-lined, radio-quiet AGN, the so-called quasar-II
population, are another potential active galaxy source of confusion for
one-lined sources, though they remain largely unidentified in
observational surveys.

\subsection{Asymmetric Line Profile}

Star formation at low and high redshift is associated with large-scale
outflows.  Resonant scattering processes in the outflowing gas will
trap \lya\ photons short-ward of the systemic velocity of the emission
and potentially destroy them through dust absorption.  This results in
a P-Cygni profile for the \lya\ line.  At the low signal-to-noise ratio
observations typically obtained on objects at $z \simgt 5$, this causes
an asymmetric \lya\ line profile with a broad red wing and a steep
cut-off on the blue wing.  This asymmetric profile is seen in many of
the confirmed $z > 5$ \lya-emitting sources \markcite{Dey:98}(\eg Dey {et~al.} 1998)
though similar asymmetries might be mimicked by low signal-to-noise
ratio observations of \oxytwo\ in the low-density limit
([\ion{O}{2}]$\lambda$3726\AA\ / [\ion{O}{2}]$\lambda$3729\AA\ $ =
0.7$).

Is an asymmetric line profile a necessary condition for high-redshift
\lya, or merely a sufficient condition?  In addition to the local
star-forming galaxies with (1) broad damped \lya\ absorption centered
at the wavelength corresponding to the redshift of the \ion{H}{2}
emitting gas and (2) galaxies with \lya\ emission marked by blueshifted
absorption features, \markcite{Kunth:98b}Kunth {et~al.} (1999) notes a third morphology of
\lya\ line that is occasionally observed in the local Universe:  (3)
galaxies showing `pure' \lya\ emission, \ie galaxies which show no
\lya\ absorption whatsoever.  \markcite{Terlevich:93}Terlevich {et~al.} (1993) present {\it IUE}
spectra of two examples of `pure' emitters:  C0840+1201 and
T1247$-$232, both of which are extremely low-metallicity \ion{H}{2}
galaxies.  \markcite{Thuan:97a}Thuan \& Izotov (1997) present a high signal-to-noise ratio {\it
HST} spectrum of the latter galaxy, noting that with $Z = Z_\odot /
23$, it is the lowest metallicity local star-forming galaxy showing
\lya\ in emission.  At high signal-to-noise ratio, the emission line
shows multiple superposed narrow absorption features, bringing into
question the `pure' designation.


Observations of local star-forming galaxies have two implications for
studies of high-redshift \lya\ emission.  First, they provide a natural
explanation for asymmetric profiles which seem to characterize
high-redshift \lya, but also imply that although the asymmetric profile
may be a sufficient condition for identification of a strong line with
\lya, it is not a necessary one.  Second, if \lya\ emission is
primarily a function of kinematics and perhaps evolutionary phase of a
starburst as suggested by the scenario of \markcite{TenorioTagle:99}Tenorio-Tagle {et~al.} (1999),
attempts to derive the comoving star-formation rate at high redshifts
from \lya\ emission will require substantial and uncertain assumptions
regarding the relation of observed \lya\ properties to the intrinsic
star-formation rate.

\subsection{Continuum Decrements}

Actively star-forming galaxies should, in the absence of dust absorption,
have blue continua,  nearly flat in $f_\nu$, at rest-frame ultraviolet
wavelengths long-ward of \lya.  This radiation derives from the hot,
massive, short-lived stars and can therefore be used as indicator of
the instantaneous star formation rate of a galaxy modulo uncertainties
in dust absorption, age, metallicity, and stellar initial mass function.
Short-ward of the Lyman limit (912 \AA), the spectral energy distributions
should drop steeply.  This is due both to photospheric absorption in
the UV-emitting stars themselves, as well as photoelectric absorption by
neutral hydrogen along the line of sight to the galaxy.  Between
the Lyman limit and \lya, photoelectric absorption from neutral hydrogen
in the intergalactic medium attenuates the emitted spectrum.  

In terms of emission line surveys, this discontinuity serves as a
useful foil for identifying observed lines with high-redshift \lya,
provided the data are sufficiently sensitive to detect continuum.
Objects which show evidence of a flat spectral energy distribution
across an emission line can immediately be ruled out as distant galaxy
candidates.  The presence of a discontinuity, however, is {\em not
sufficient} for classifying the observed emission line with \lya,
unless the amplitude is extreme (see below).  In particular, \oxytwo,
the other strong, solitary emission feature in the UV/optical spectra
of star-forming galaxies, lies short-ward of the Balmer and 4000
\AA\ breaks.  The former is strongest in young systems dominated by an
A-star population, while the latter arises from metal-line blanketing
(predominantly \ion{Fe}{2}) in late-type stars.  At high
signal-to-noise ratio, the morphology of the break can be used to
distinguish the \lya\ forest from the Balmer break from the 4000
\AA\ break.  At low signal-to-noise ratio, however, the continuum
decrement across \oxytwo\ can mimic that across high-redshift \lya.
For example, \markcite{Dey:99a}Dey {et~al.} (1999) report an optical
spectrogram of the extremely red object ERO~J164502+4626.4
\markcite{Hu:94}(HR10 in Hu \& Ridgway 1994), showing a single, strong
emission feature at 9090.6 \AA\ and a drop in the continuum level by a
factor of $\approx 3$ across the emission line.  The optical data alone
is suggestive of \lya\ emission at $z = 6.48$.  However, near-infrared
images show the source to be extremely red, with a steeper spectral
energy distribution than expected for a high-redshift, star-forming
galaxy.  Furthermore, a near-infrared spectrum shows a second emission
at 1.603$\mu$m, solidying the emission line identifications as
\oxytwo\ and H$\alpha$ at $z = 1.44$.  Extremely dusty,
moderate-redshift galaxies such as ERO~J164502+4626.4 may be
misidentified as high-redshift \lya-emitters by {\em solely} optical
surveys.

The amplitude of a continuum discontinuity may be used as a tool for
distinguishing the identification of that discontinuity.  In order of
decreasing wavelength, discontinuities are commonly observed in
UV/optical spectra of galaxies at rest wavelengths of 4000
\AA\ [$D(4000)$], 2900 \AA\ [$B(2900)$], 2640 \AA\ [$B(2640)$], 1216
\AA\ (\lya), and 912 \AA\ (the Lyman limit).  The hydrogen
discontinuities derive from associated and foreground absorption and
thus have no theoretical maximum.  The longer rest-wavelength
discontinuities derive from metal absorption in the stars and galaxies
and are thus dependent on the age and metallicity of the galaxy
\markcite{Fanelli:92}(\cf Fanelli {et~al.} 1992).  The largest measured
values of $D(4000)$ are $\sim 2.6$ \markcite{Hamilton:85,
Dressler:90}(Hamilton 1985; Dressler \& Gunn 1990), while {\it IUE}
spectra of main-sequence stars exhibit $B(2900) \simlt 3$ and $B(2640)
\simlt 3$ \markcite{Spinrad:97}(Spinrad {et~al.} 1997).  The example of
ERO~J164502+4626.4 illustrates the utility of break amplitudes for
redshift identifications.  The amplitude of the continuum break in
ERO~J164502+4626.4 is $\approx 3$ across an emission line at 9090.6
\AA.  Were the emission line \lya, the implied redshift would be $z =
6.48$.  Models and measurements of the strength of the \lya\ forest
imply decrements $\simgt 10$ at $z \simgt 6$ \markcite{Madau:95,
Stern:99e}(Madau 1995; Stern \& Spinrad 1999), incompatible with the
observed decrement and arguing for a lower redshift line
identification.

\section{Conclusions}

We report the serendipitous discovery of two faint galaxies with high
equivalent width emission lines at long wavelength ($\lambda \approx
9190$ \AA).  For one source, galaxy B, faint blue continuum and a weak
secondary line emission feature implies the source is at $z = 1.466$.
The spatial proximity and similar emission wavelength of object A
persuasively argues that this is an unusual \oxytwo-emitter at $z =
1.464$.  It had been thought that serendipitous one-lined sources with
rest-frame equivalent widths larger than a few $\times$ 100 \AA\ are
exclusively identified with high-redshift \lya.  Our observations have
shown that this is demonstrably not the case.  Both sources are unlike
local star-forming galaxies, and we suggest that the observations are
most consistent with the discovery of a moderate-redshift active system.

Several programs are currently underway to find high-redshift \lya\
emitters using a combination of narrow- and broad-band imaging
\markcite{Thommes:98, Hu:98}(\eg Thommes {et~al.} 1998; Hu {et~al.}
1998) and serendipitous long slit surveys \markcite{Manning:00}(\eg
Manning {et~al.} 2000).  The moderate-redshift system discussed herein
has serious implications for those surveys.  In particular we expect
the line-emission surveys to uncover a population of strong line
emitters and we emphasize that comparison of this sample with
magnitude-limited surveys could be misleading.

We discuss various criteria which can be used to assess whether a
solitary, high-equivalent width emission feature is associated with
high-redshift \lya.  In Table~2 we consider how recently reported $z >
5$ spectroscopic candidates fare with respect to these criteria.  Some
sources are reliably confirmed at $z > 5$, while others are confirmed
at low-redshift and yet others remain ambiguous with the current data.
We conclude that some criteria are necessary, but not sufficient, to
conclude that a source is at $z > 5$, such as a continuum decrement
across the emission feature, while other criteria are sufficient, but
not necessary, such as an asymmetric line profile.  We suggest that
multiple criteria are necessary to convincingly demonstrate that a
single-lined source is at high-redshift.

\acknowledgments

We thank Alex Filippenko, Doug Leonard, and Aaron Barth for obtaining
the July 1997 observations, and Tom Broadhurst and Brenda Frye for the
follow-up observations during August 1997.  We are indebted to the
expertise of the staff of Keck Observatory for their help in obtaining
the data presented herein, and to the efforts of Bev Oke and Judy Cohen
in designing, building, and supporting LRIS.  The work presented here
has been aided by discussions with Mike Liu, Curtis Manning, Gordon
Squires, and Chuck Steidel.  We are also grateful to Trinh Thuan for
sharing the {\it HST}/GHRS spectrum of T1214$-$277 and to Carlos
De~Breuck and Adam Stanford for carefully reading the manuscript.  This
work has been supported by the following grants: IGPP/LLNL 98-AP017
(DS), NICMOS/IDT grant NAG~5-3042 (AJB), NSF grant AST~95-28536 (HS),
and NASA HF-01089.01-97A (AD).  A preliminary version of this work was
presented at the January 1999 AAS meeting in Austin under the title
``Cosmological One-Liners:  A Cautionary Tale''
\markcite{Stern:99d}(Stern {et~al.} 1999a).



\eject

\end{document}